\documentclass[10pt,notitlepage,onecolumn]{article}

\usepackage{amsmath}
\usepackage{amssymb}
\usepackage{bm}
\usepackage{graphicx}
\usepackage{amsbsy}
\usepackage{color}

\begin{document}


\begin{flushleft}
{\Large
\textbf\newline{Intriguing yet simple skewness - kurtosis  relation in economic and demographic  data distributions; pointing to preferential attachment processes} 
}
\newline
\\
  Marcel Ausloos \textsuperscript{1,2,*} and Roy Cerqueti \textsuperscript{3}
 \\
\bigskip
 \textbf{$^1$} School of Business, University of Leicester, University Road. Leicester, LE1 7RH, United Kingdom. Email: ma683@le.ac.uk\\
 \textbf{$^2$}  GRAPES -- Group of Researchers for Applications of Physics in Economy and Sociology. Rue de la Belle Jardini\`ere 483, B-4031, Angleur, Belgium. \\Email: marcel.ausloos@ulg.ac.be
\\
 \textbf{$^3$} Department of Economics and Law, University of Macerata, Via Crescimbeni 20, I-62100, Macerata, Italy.
 Tel.: +39 0733 258 3246; Fax: +39 0733 258 3205.\\Email: roy.cerqueti@unimc.it.
 \\
\bigskip




\end{flushleft}


\begin{abstract}
In this paper,  we propose that  relations between high order moments of data distributions,  for example  between the skewness (S) and kurtosis (K),  allow to point to theoretical models with understandable  structural parameters.
The illustrative data concerns  two cases:  (i) the distribution of  income taxes and (ii) that of  inhabitants, after aggregation over each city in each province of Italy in 2011.
Moreover, from the rank-size relationship, for either S or K, in both cases,  it is shown that one  obtains the parameters of the underlying (hypothetical) modeling distribution: in the present cases, the 2-parameter Beta function, - itself related to the Yule-Simon distribution function, whence suggesting a  growth model based on the preferential attachment process.
\end{abstract}


\section{Introduction}\label{Introduction}

Characteristics of distributions of variables  is a never ending
subject of investigations in many fields of economic research. There
is much work for example on testing conditional convergence in
variance and skewness for estimating (multivariate or not) normality
(Mardia,  1970; L\"utkepohl and Theilen,  1991; Nath, 1996; Bai and
Ng, 2005; Egger and Pfaffermayr,  2009; Huber  and Pfaffermayr,
2010),  or on  improving goodness-of-fit in regressions (Dufour et
al., 2003).   In fact, drawing inference on the parameters of
(regression or agent based) models is a basic statistical problem
which  fortunately may provide  interesting discoveries  (Sahota,
1978; Lin  and Lui,  1992; Richardson  and Smith,  1993), not only
on strict economic problems,  but also in related sociological or
demographic ones (Eeckhout,  2004;  Kaitila, 2014; Vitanov and
Ausloos, 2015; Cerqueti and Ausloos, 2015a, 2015b, 2015c).

There has been already much interesting work on recurrence relations
between high moments ($\mu_i$, $i \ge 3$) of order statistics
distributions (Arnold et al., 1992; Malik et al., 1988). Knowledge
of such moments are of interest  for drawing inference about the
scaling parameters.
Balakrishnan et al.  (1988) 
have reviewed many recurrence  relations and identities for several
continuous distributions.
Pertinently,  in view of the following, let
us point to Thomas and Samuel  (2008) 
analysis of  recurrence relations for the  Beta distribution
moments, - a distribution of wide application, both in its
continuous or discrete form (Johnson et al., 1995; Martinez-Mekler
et al., 2009;  Ausloos and Cerqueti, 2016a)

The  skewness and  kurtosis  are officially the third and  fourth moment, $\mu_3$ and $\mu_4$ respectively, of a distribution, where
$\mu_i$ is usually centered on the mean; $\mu_2$ is of course the variance, sometimes called $\sigma^2$.  Mathematical statistics
textbooks and  software packages   usually calculate the Fisher-Pearson coefficient of skewness and the kurtosis
  \begin{equation} \label{S}
 S= \mu_3\;/\; \mu_2^{3/2}
\end{equation}
 \begin{equation} \label{K}
 K = \mu_4\;/\; \mu_2^2.
\end{equation}
respectively, which by extension become the (commonly accepted measure of) skewness and kurtosis of the distribution, respectively;
we use such notations here below.

It is known since Pearson (1916) 
and Wilkins (1944) 
 that the kurtosis has a theoretical lower bound  related to the skewness $ K \ge a S^2+b$. Based on  several
types of experimental data, one has indeed observed that  the quadratic relationship
 \begin{equation}\label{KpS2q}
 K= p\;S^2 + q
 \end{equation}
holds as an envelope of scattered data (see references below).
In fact,   $p$ and $q$ can be empirically fitted constants,
 which might have some  interesting meaning, as pertinently shown by
 \textbf{Guszejnov et al.  (2013).} 




It has recently been discussed by \textbf{Cristelli et al. (2012)}
that a more general form of the $K-S$ relationship could be provided, i.e.
\begin{equation}\label{eqKSnu}
K = p\; S^{\nu} + q
\end{equation}
with $\nu= 4/3$. However the findings (and interpretation) have been questioned: such an exponent might be due to the data too limited size (Celikoglu and Tirnakli, 2015) 

Much of such experimental data, for which Eq. (\ref{KpS2q}) is obeyed,  pertains to magnetohydrodynamics, meteorological  and
medical data. Moreover,  most of the data where such relations are found pertain  to time series.

We provide another set of cases in which one finds $\nu=2$,  in an unusual set of data pertinent to a complex geo-sociological-economic realm.
Eq. (\ref{eqKSnu}), with $\nu\simeq 2$,  is found  to hold for
  the wealth  and population distribution of Italy cities, aggregated at the provincial level.
We tie this specific  finding to a statistical process inferring a Beta-distribution, or Polya urn dynamics, as its universal dynamics.

In Section \ref{litrev}, we very briefly recall previous experimental data analysis with similar  findings, pertinent to the present report, but obviously  in
quite 
(in scientific terms) scattered  fields of investigations, - in order to stress the originality of the present ones.


In Section \ref{ITdata}, we explain the   system complexity which we
investigated. 
Notice at once, that there is no time dependence, therefore the
cloud of points, from which other works  infer a parabola type relation between $S$ and $K$,    is here strictly "reduced" to a collapsing-like
situation, rendering the parameter values being much more precise in view of describing the dynamical process, -- if we are allowed to discriminate
between (so called) hard and soft science.
Such a Section contains also the main methodological investigation methods, results and related comments.

Section \ref{modelBeta} is devoted to the description of a
theoretical Polya urn model which is related to the developed complex system arguments. Specifically, a preferential attachment
system is introduced whose stochastic law follows a Beta distribution,  whence whose parameters could be also calibrated for defining
skewness and kurtosis of a set of data.

In Section \ref{conclusion}, we offer some conclusive remarks and provide also suggestions for future research directions.

Some Tables containing descriptive statistics of the data and the disaggregation of them at a provincial level are reported in the 
Appendix. 
%
%
%
%
%
%
\section{Literature review}\label{litrev}
As mentioned in the introduction, there are several reports pointing
to the veracity of  Eq. (\ref{eqKSnu}), with $\nu=2$,   in various
research fields: the greatest occurrence is in turbulence, among the
most recent see  in magneto-hydrodynamics
 (Labit et al., 2007; \textbf{Guszejnov et al., 2013;} Bergsaker et al., 2015)
and in atmospheric physics (see  Sura  and Sardeshmukh  (2008)).
In this respect, refer also to Alberghi et al. (2002), where the
authors discuss the parameters conditions to be satisfied in the
context of air vertical velocity in the atmospheric boundary layer
for having a certain relation between skewness and kurtosis. In
these cases, in order to have different points in the $S-K$ plane,
the authors usually repeat the experiment or evaluate the moments on
different time windows of the same series. \textbf{For further pertinent
references, e.g. see
 Mole and Clarke (1995) and Sattin et al. (2009).}

Related to the above-mentioned papers, there is a study in the context of geophysics of oil production time series forecasting (Frausto-Sol{\'\i}s et al., 2015).

\textbf{Cristelli et al. (2012)} 
analysed the relation between skewness and kurtosis for earthquakes   and  daily price
returns (on the S\&P500)  and identified two power-law regimes of non-Gaussianity,  on the kurtosis versus skewness plots, but
 Celikoglu and    Tirnakli  (2015) 
demonstrated that the proposed "universal" relation between skewness and kurtosis,  in fact   is not universal and originates only due to
the small number of data points in the data sets considered.

\textbf{For completeness, on  $K-S$  relation consideration in the
financial domain, let us mention related work on   $S^2-K$ bounds
for unimodal distributions by Klaassen et al. (2000), and more
recently  by McDonald et al., (2013) and  Kerman  and McDonald,
(2015) with a discussion about modeling  some "popular income
distributions"  with exponential generalized beta functions. }

Other cases where a simple  $K-S$  relation is found are  in surface
roughness analysis   (Isoda and Kawashima,  2013; Sharifi-viand et
al., 2014).
 It is also worth mentioning the contributions in medical and biological fields; they pertain to
 fluctuation responses in the visual cortex (Medina and Diaz, 2016a, 2016b, 2016c) 
 or  ventricular fibrillation (Gonzalez-Gonzalez et al., 2013);   an apparently comprehensive review about "biological and psychological aspects of the $K$-$S$
 can be found in  Cain et al. (2016). 

One might put in parallel to the above a paper reporting a 4/3 relation between skewness and kurtosis of aesthetic score
distributions in a  photo aesthetics dataset, generated from an online voting survey (Park and  Zhang 2015).  


Thus, it maybe observed that  most of the data pertains to time
series analysis,   many to human reaction time, and a few \textbf{
to} various (laboratory or not) produced crystalline samples.
However, to the best of our knowledge, not many observations of a
peculiar relationship between $K$ and $S$ seem to have been reported
on socio-demography aspects.

\section{Italy economic and demographic data}\label{ITdata}

A statistical assessment of regional wealth inequalities over Italy (IT) has been previously provided based on aggregated tax income size data
(Mir et al., 2014; Cerqueti and Ausloos, 2015a, 2015b, 2015c;  Ausloos and Cerqueti, 2016b). 


Let it be known that IT is nowadays (since 2010) made up of 8092
cities distributed over 110 provinces. To provide some better
understanding of the paper aims and results, the IT administrative
structure  can be  briefly described as follows. Italy is clustered
in 20 non-overlapping regions, and each region contains one or more
provinces, which in turn are composed by cities along with their
territories ($comuni$). Thus, each city belongs to only one
province, and each province is contained in only one region.

The Gross Domestic Product (GDP) in Italy was worth about 2276
billion USD   in 2011.  The population of Italy \textbf{fell}
"slightly" below 60 millions, then.

The economic data was obtained from (and by) the Research Center of the Italian Ministry of Economics and Finance (MEF).
The  population data source  is the Italian Institute of Statistics (ISTAT). In particular, data on the population are extracted from the elaborations of the 15th Italian Census, performed by ISTAT in
2011. We have disaggregated contributions at a municipal level for 2011, in order to obtain the  aggregated  tax income (ATI),  $ATI_c$
and the number of inhabitants $N_{inhab,c}$, for each city $c$.

%

   \begin{table} \begin{center}
\begin{tabular}[t]{|c|cc||cc|}
  \hline
& \multicolumn{2}{|c|}{ $ATI_{c,p}$}&  \multicolumn{2}{|c|}{ $N_{inhab,c,p}$}  \\ \hline
 &  $S $ &  $K $ & $S $ & $K $  \\\hline
Min.    &   0.58791 &   -1.3075 & 0.76111   &   -0.93186    \\
Max.    &   17.092  &   296.36  & 16.951    &   292.99  \\
Sum &   655.56  &   4982.1  & 594.05    &   4291.0  \\
$N_p$   &   110     &   110     & 110   &   110     \\
Mean ($\mu$)    &   5.9596  &   45.292  & 5.4005    &   39.009  \\
Median ($m$)  &   5.7346  &   35.458  & 5.0314    &   27.925  \\
RMS     &   6.5453  &   62.251  & 6.0163    &   56.184  \\
St. Dev.  ($\sigma$)  &   2.7188  &   42.902  & 2.6636    &   40.619  \\
Variance    &   7.3920  &   1840.6  & 7.0948    &   1649.9  \\
Std Err.    &   0.25923 &   4.0905  & 0.25396   &   3.8729  \\
Skewn.  &   0.87472 &   2.5531  & 1.0765    &   2.9083  \\
Kurt.   &   1.6629  &   10.238  & 2.2291    &   13.121    \\\hline
 $\mu/\sigma$       &   2.1920  &   1.0557      &   2.0275  &       0.9604          \\
 3$(\mu-m)/\sigma$      &   0.2483  &   0.6877  &   0.4157  &     0.8186        \\
 $\rho$ &   -0.1234     &   3.2129      &   0.0840  & 4.282 \\\hline
 $\mu-2\sigma$ &   0.5219     &   -40.511      &   0.07328  & -42.2298 \\
$\mu+2\sigma$ &   11.397     &   131.09    &   10.728  & 120.248 \\
\hline
\end{tabular}
   \caption{Summary of
(rounded) statistical characteristics  for  the distribution of  $S$
and $K$ for $ATI$ of cities   in IT provinces and  for the
distribution of  the number of inhabitants in IT cities in the
various ($N_p=110$) provinces. \textbf{The last two lines will be
useful in the assessment of the outliers, see below.}
}\label{TablestatSKSK}
\end{center} \end{table}

   \begin{table} \begin{center}
\begin{tabular}[t]{|c|c||c|c|}
  \hline
Equations & &   $ATI_{c,p}$ &       $N_{inhab,c,p}$ \\\hline
&$\nu$ (theoretical)&    2   &       2   \\
Eq. (\ref{KpS2q}) & $p$         &   1.048$\pm$0.008 &   1.052$\pm$0.010\\
&$q$         &   0.415$\pm$0.499  &0.921$\pm$0.518   \\
&$R^2$       &0.993  &0.991      \\ \hline
&$\nu$ (empirical)   &   1.912$\pm$0.026 & 1.894$\pm$0.028           \\
Eq. (\ref{eqKSnu}) & $p$         &   1.324$\pm$0.092 & 1.389$\pm$0.103   \\
&$q$         &   -2.065$\pm$0.900    & -1.763$\pm$0.898  \\
&$R^2$       & 0.994 & 0.992     \\ \hline
&$\rho$  &    5.7224   &      5.9116\\
Eqs. (\ref{Beta})-(\ref{fx}) & $a $ &  0.7556    & 0.8493  \\
&$b $  &4.9668  &    5.0623    \\
 \hline
\end{tabular}
\caption{Summary of (rounded) parameters for the quadratic relation
between $K$ and $S$: Eq. (\ref{KpS2q}); for the power law relation
between $K$ and $S$: Eq. (\ref{eqKSnu}); for the Beta-distribution
model: \textbf{Eqs. (\ref{Beta})-(\ref{fx})} -- about $ATI$ of IT
cities ($N_c=8092$) and number of inhabitants in the $N_p=110$ IT
provinces in 2011. For the Beta case, we   report  a
 graph of the Beta \textbf{cumulative distribution} functions
with the relevant calibrated parameters, in  Figures
\ref{fig:BetaATI} and \ref{fig:BetaNinhab}. Some related comments
can be found in Subsection \ref{sub:beta}. The value of the
auxiliary parameter $\rho$ is reported for the sake of completeness.
\textbf{The $\pm$ refers to the standard error value.}}
\label{pqTableparameters}
\end{center} \end{table}

  \begin{table} \begin{center}
\begin{tabular}[t]{|c||c|c||c|c||c|}
  \hline
Eq.(\ref{Lavalette4up})& \multicolumn{2}{|c||}{ $ATI_{c,p}$}&  \multicolumn{2}{|c||}{ $N_{inhab,c,p}$} &Eq.(\ref{Beta}) \\ \hline
&   $K$ &       $S$ &   $K$ &       $S$ &  \\\hline
 $\kappa_4$         &   3.1426$\pm$0.525    &   1.3666$\pm$0.075 &1.3447$\pm$0.244&1.7307$\pm$0.993&\\
$\gamma_4$      &   0.2884$\pm$0.009  &0.1248$\pm$0.005&0.2865$\pm$0.009&0.1692$\pm$0.005&$b-1$ \\
$\xi_4$         &   0.8853$\pm$0.033    &   0.4816$\pm$0.010 &1.0434$\pm$0.036& 0.4378$\pm$0.010&$a-1$\\
$\psi_4$        &   0.2649$\pm$0.031     &0.1225$\pm$0.024& 0.1897$\pm$0.022& 0.2812$\pm$0.039& \\\hline
$R^2$       &0.9947 &0.9945     &0.9956&0.9951&\\ \hline
\end{tabular}
   \caption{Summary of
(rounded) parameters for the fits to Eq. (\ref{Lavalette4up}) based
on Beta-distribution model about $ATI$ of IT cities    in 2011 and
number of inhabitants in the  $N_p=110$ IT provinces from the
2007-2011 census; the correspondence between the exponent parameters
and the Beta function parameters is given for completeness.
}\label{TableLav4parameters}
\end{center} \end{table}

  \subsection{$K$-$S$ relation  analysis}\label{analysis}
What we care about is the distributions of the skewness and of the
kurtosis: these are the distributions of interest.

The interesting Table is Table \ref{TablestatSKSK}. It is seen that the corresponding (in some sense "average" $S$ and $K$ are positive
and not small. The most relevant point seems to be the existence of (2) negative $K$ values:  -1.30750   for BT (Barletta-Andria-Trani, Apulia region)  in the $ATI$ case
and   -0.9319  for BT and  -0.7511    for RG  (Ragusa, Sicily) for the $N_{inhab}$.

On the other extreme TO (Torino province in Piedmont region) has the largest $K$ and $S$ for both cases; for information TO contains the
largest number of cities (315).

 The obtained corner results for these specific provinces are in line with   historical and empirical "evidence". In
fact, Torino represents the core of the industrial production of
Italy, being the headquarter of FIAT. Thus, the related province has
an \textbf{unequal}  distribution of richness and population, with
an asymmetry of positive type. It is also expected that the tails of
the distribution are heavy, being Torino (the city) one of the
largest cities in Italy, -- in terms of number of inhabitants and
ATI, -- and since the same province contains some of the smallest
cities of the Country.

For what concerns BT and RG, they are two of the less populated and poor provinces in Italy, with a large part of small cities. Hence, a
platycurtic distribution for the $ATI$ and $N_{inhab}$ is what everyone with a fair level of knowledge of the Italian reality should expect.

\begin{figure} 
    \includegraphics[width=1.250\textwidth]{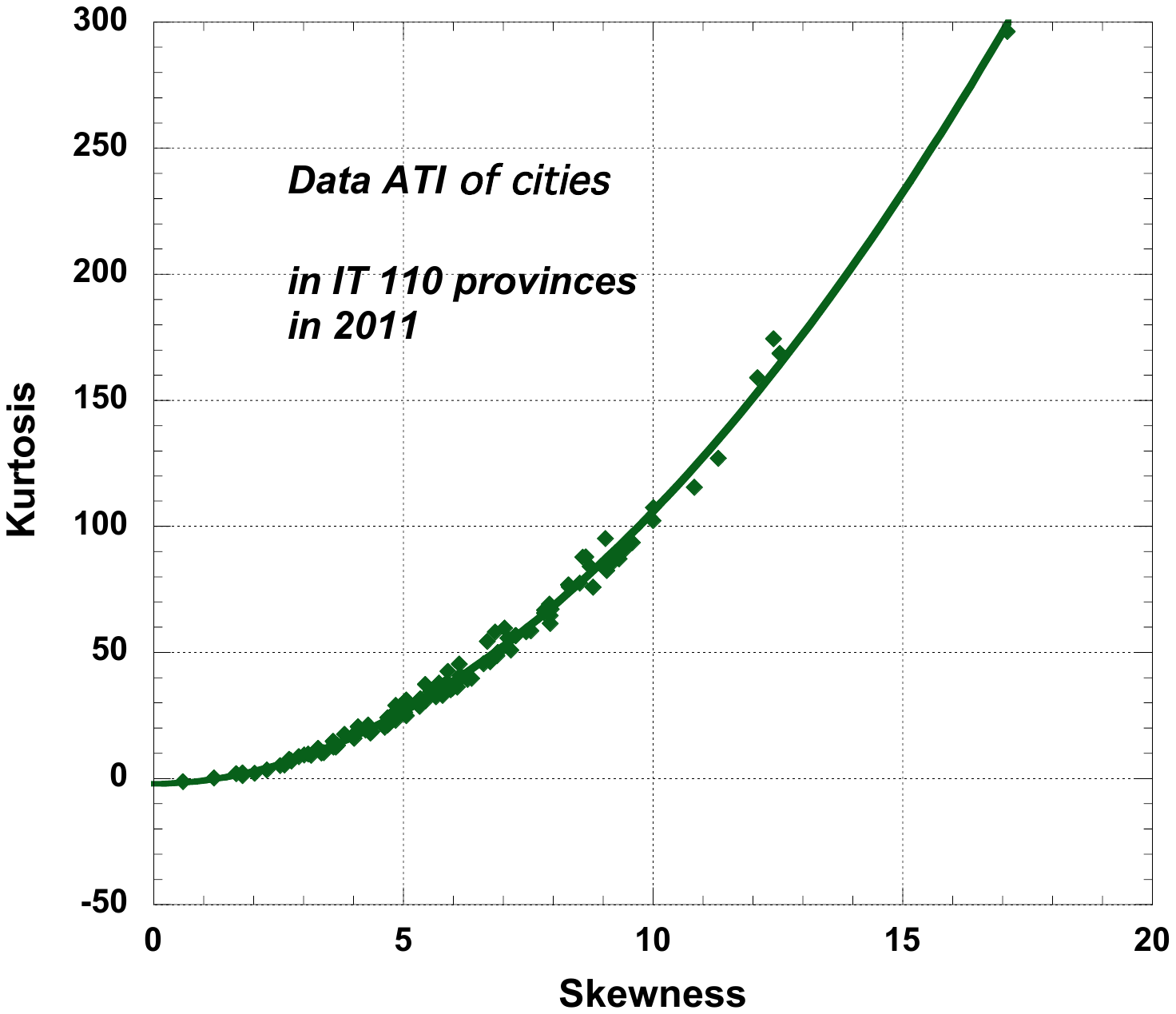}
    \caption{$K$-$S$ relationship, Eq. (\ref{eqKSnu}),
for the distribution of ATI aggregated over cities in the IT 110
provinces  in 2011; $\nu\simeq 1.91$.}\label{fig:KS2ITATI}
\end{figure}

The $K$-$S$ relationship for the distribution of ATI of cities in the 110 provinces  is given in Fig. \ref{fig:KS2ITATI}

\begin{figure} 
    \includegraphics[width=1.25\textwidth] {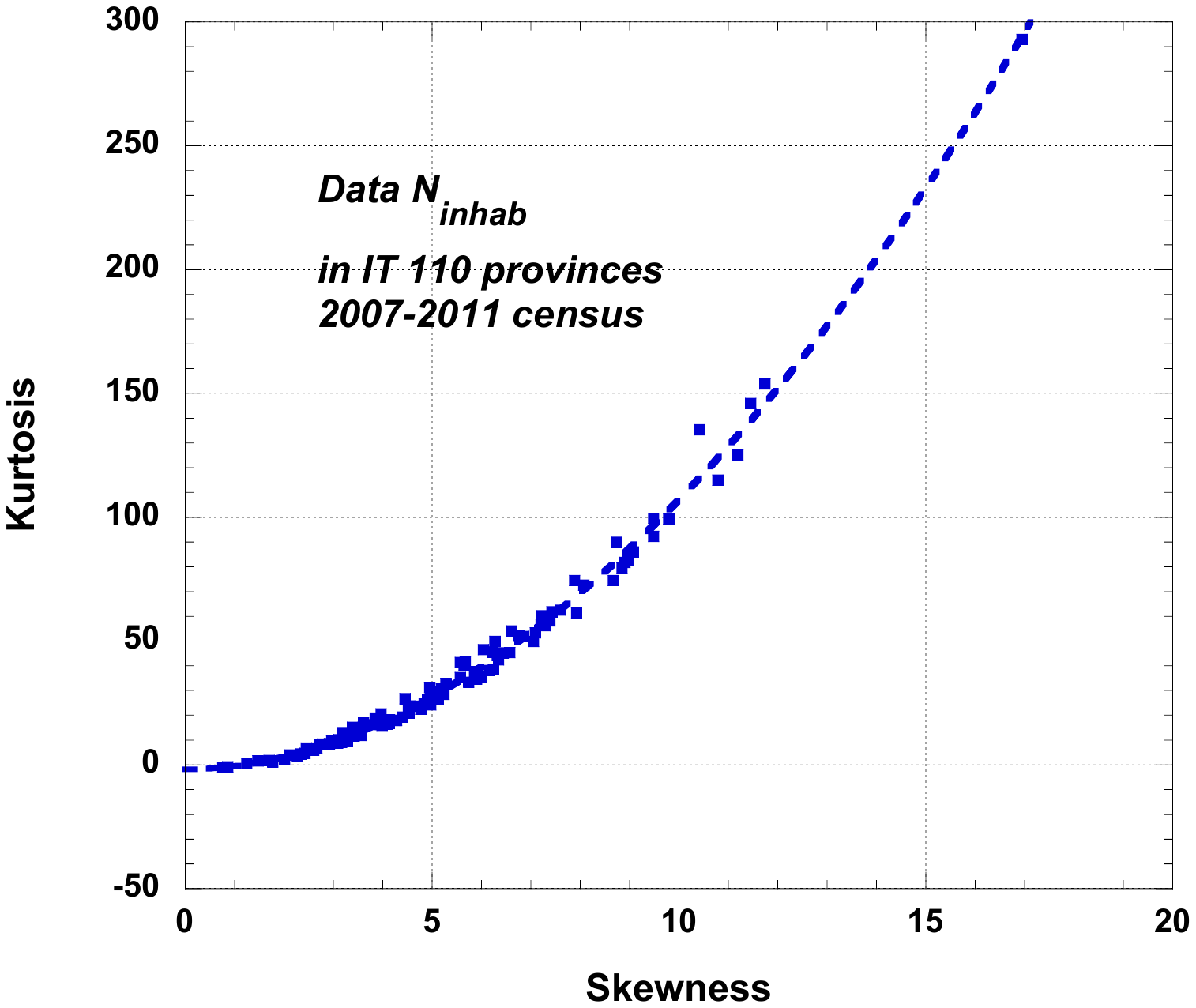}
\caption{$K$-$S$ relationship, Eq. (\ref{eqKSnu}), for the
distribution of the number of inhabitants of the IT  cities in the
110 provinces  according to  the 2011 Census; $\nu\simeq
1.89$.}\label{fig:KS2ITNinhab}
\end{figure}

The $K$-$S$ relationship for the distribution of the number of inhabitants  in   cities in the 110 provinces is shown in Fig. \ref{fig:KS2ITNinhab}. The relation is pretty smooth in both cases,
and recall those found in magnetohydrodynamics and other studies of time dependent systems;  see pertinent references in Sect.\ref{litrev}.

Thereafter,  in accord with previous literature, we try two fits:
%
%
%
a polynomial of degree 2 but without linear term,  i.e. Eq.(\ref{KpS2q}), or the pseudo parabolic polynomial-like  form
Eq.(\ref{eqKSnu}).

The parameters $p$ and $q$ are given in Table \ref{pqTableparameters}, together with the corresponding regression
coefficient. To leave $\nu$ as a free fit parameter is seen not to
be a drastic improvement. Thus, one can expect, since $\nu \simeq 2$
that a simple interpretation or modelisation based on well
established statistical distribution is in order; see Sect.
\ref{modelBeta}.

It can be noticed that $p\simeq 1$, but $q$ is negative, since there is a negative kurtosis for the distributions in a couple of
provinces (see Table \ref{TablestatSKSK}).

\subsection{Rank-size analysis}\label{ranksizeanal}

\textbf{The above findings remind us that there is a general
relationship between skewness and kurtosis within Pearson's
distribution system.  Therefore, in order to pursue toward some
understanding of this sort of  $K-S$ relation, we propose to develop
a complementary analysis.
Instead of considering the $K$ and $S$ values as belonging to some
continuous distribution, we are using a method, the rank-size
analysis method,  which allows to study  a distribution of
"quantities" when the orders of magnitudes can be rather different,
and when the values have some imprecise error bar, - as it always
occurs in such economic and sociological surveys. In such a
methodology,  the ($K$ and $S$ here) values are supposed to belong
to discrete distributions which  are regularly sampled. Thus, we
write the  $S$ and $K$  data  in an ascending size (regular) order,
independently of each other, for the ATI and the number of
inhabitants, respectively, i.e. giving the rank $r=1$ to the lowest
$S$ and to the lowest $K$ values, etc.}

The most simple rank-size law is thought to be a power law of the
rank $r$, - leading to the Zipf plot, $y\sim r^{-\alpha}$. It is
often modified for including  an upper tail cut-off as through the
Yule-Simon law,
 \begin{equation} \label{PWLwithcutoff}
 y(r)= d \;r^{-\alpha} \; e^{-\lambda r}.
\end{equation}
In order to take into account a possible change of  curvature in the data, if
some falling off  seems to occur  visually, at the highest ranks, 
 Eq. (\ref{PWLwithcutoff}) can be then written  (Ausloos, 2014a, 2014b; Ausloos and Cerqueti,  2016a) as
\begin{equation} \label{Lavalette3a}
y_3(r)= \kappa_3\;    r^{- \gamma}\; (N-r+1)^{-\xi}  ,
\end{equation}
where $+1$ is introduced in order to avoid a singular point in the fit at the highest rank $r_M = N$, if  $\xi\simeq 0$ . This also emphasizes that an upper tail  toward infinity is rather meaningless, since the upper rank $r_M$ is necessarily finite.


In view of taking into account a better fit at low and high rank,
one can further generalize Eq. (\ref{Lavalette3a}) to a five
parameter free equation  (Ausloos and Cerqueti,  2016a): 
\begin{equation} \label{Lavalette5}
y_5(r)= \kappa_5\;   (r+\Phi)^{- \gamma} \;(N+1-r+\Psi)^{-\xi}  ,
\end{equation}
where the parameter $\Phi$ tis reminiscent of Mandelbrot's generalization of Zipf's law at low rank, 
while $\Psi$ allows some flexibility at the highest rank, -- where
usually the error bar on the data can be rather influential in
defining $r$. The shape of the curve in Eq. (\ref{Lavalette5}) is
sensitive to the variations of $\Phi$ and $\Psi$  (Ausloos and Cerqueti,  2016a).

  Here, neglecting any low rank   free parameter ($\Phi$) of dubious origin, but still
allowing for some flexibility on the upper rank value divergence,
we approximate Eq. (\ref{Lavalette5}) by
\begin{equation} \label{Lavalette4up}
Y_4(r)=
\kappa_4 \; r^{\xi_4}  \;   (N-r+\psi_4)^{-\gamma_4}
\end{equation}

\textbf{One is allowed to imagine that a Generalized Discrete Beta
function, like Eq.(\ref{Lavalette4up}), reminds the reader of the
Pearson Type I distribution, supported in the  relevant rank
interval $[0,N]$.} In so doing, the corresponding best  fits of the
various $S$ and $K$ ranked data  can be found
 for the  distributions of
city $ATI$  for 2011 and of the number of inhabitants of the IT
cities   according to  the 2011 census, distributed over the 110
provinces, respectively.

\begin{figure} 
    \includegraphics[width=1.250\textwidth]{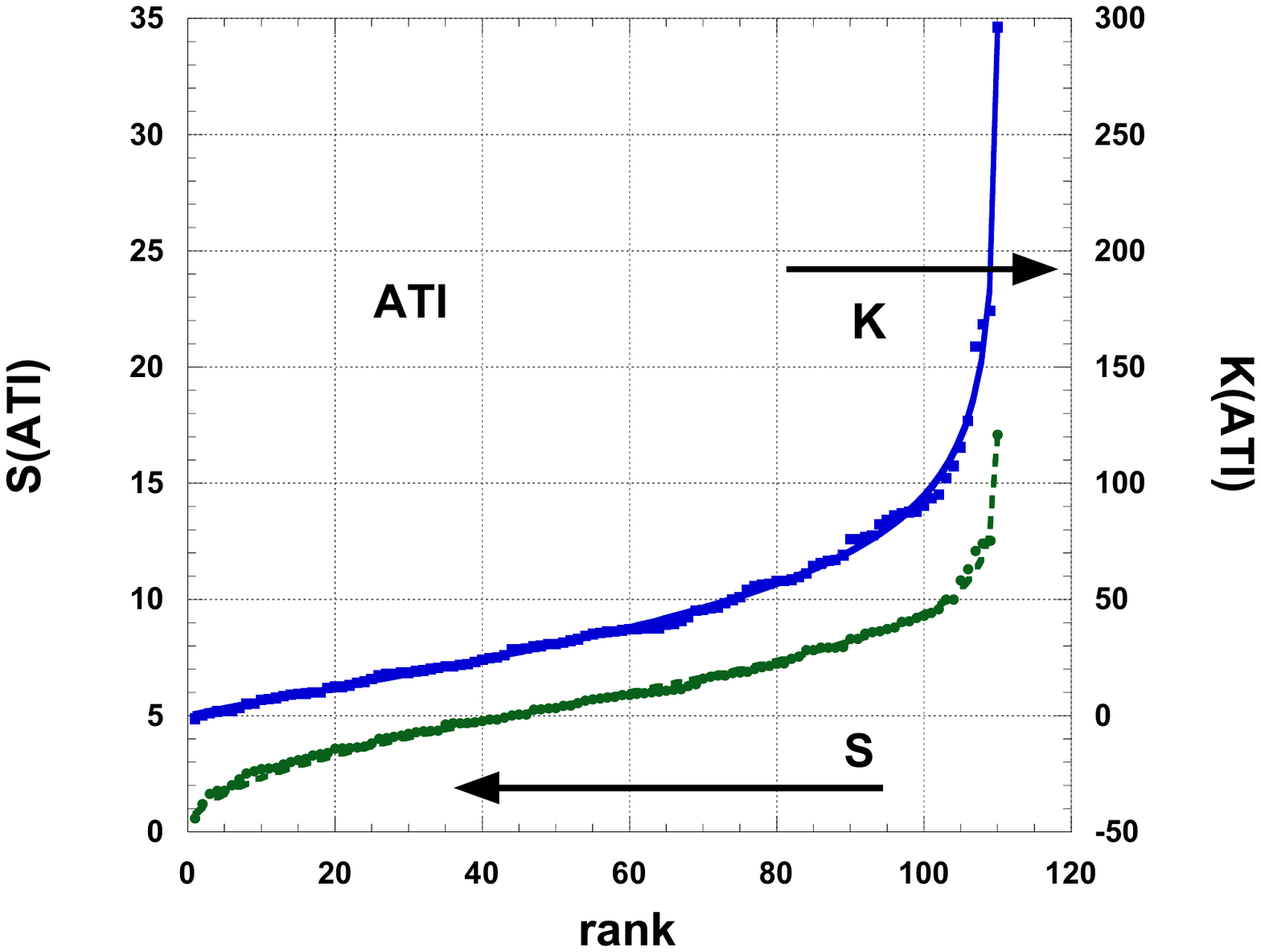}
    \caption{Rank-size  relation for $K$ and $S$, with fits with Eq. (\ref{Lavalette4up}),
for the distribution of ATI aggregated over cities in the IT 110
provinces  in 2011; fit parameters in  Table \ref{TableLav4parameters}.
}\label{fig:ranksizeKSITATI}
\end{figure}

\begin{figure} 
    \includegraphics[width=1.25\textwidth] {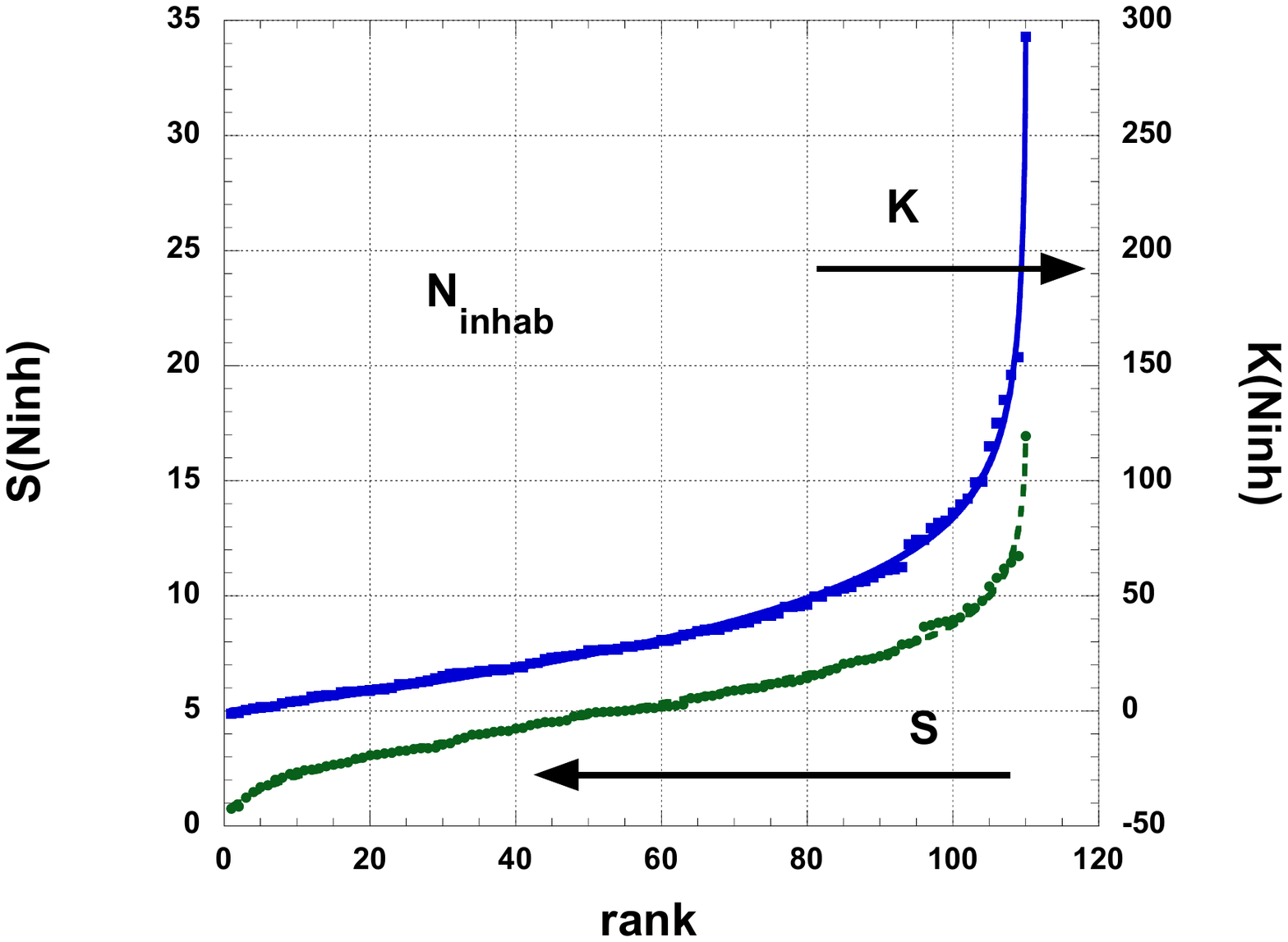}
\caption{Rank-size  relation for $K$ and $S$, with Eq. (\ref{Lavalette4up}) best fits, for the
distribution of the number of inhabitants of the IT  cities in the
110 provinces  according to  the 2011 census; fit parameters in
Table \ref{TableLav4parameters}.}\label{fig:ranksizeKSITNinhab}
\end{figure}

\textbf{Fig. \ref{fig:ranksizeKSITATI}   displays the rank-size
relation for $K$ and $S$ for the distribution of ATI aggregated over
cities in the IT 110 provinces in 2011,  and the best fits by Eq.
(\ref{Lavalette4up}). In the same spirit, Fig.
\ref{fig:ranksizeKSITNinhab} gives  the rank--size  relation for $K$
and $S$, with fits by Eq. (\ref{Lavalette4up}), for the distribution
of the number of inhabitants aggregated over cities in the IT 110
provinces  according to the 2011 census. The  fit parameters are
found  in Table \ref{TableLav4parameters}.  The  values (and error
bars) on  the $ \xi_4$ and $ \gamma_4$, together  with the  value of
the regression coefficient, $R^2$,  are quite convincing of the
existence of an inflection point in the  rank-size data.
Nevertheless, observe the variety of values for  $ \xi_4$ and $
\gamma_4$.}

 \section{Polya Urn Modelization}\label{modelBeta}

\subsection{Preferential attachment}
Our argument for suggesting a model stems from the historical view
that cities do not appear nor grow stochastically, Moreover,  there
is a postulate on demography that ghettos form according to peer
status, in particular, due to the wealth of the population: rich and
poor group themselves in clusters. In so doing, the number of
inhabitants is somewhat related to the wealth. A similar type of
process can be imagined, {\it mutatis mutandis}, for such different
qualities: the so called  "preferential attachment process". Such a
process can be defined as a settlement procedure in urn theory,
where additional balls are added and distributed continuously to the
urns (cities, in this model) composing the system.
 The obtained model is the general Polya urn  (Mahmoud, 2009).
In our context, the rule of such an
addition follows an increasing function of the number of balls
already contained in the urns.  The settlement
formation obeys a Yule process, with a log-normal initial
distribution of the population of the settlements.

In general, such a process contemplates also the creation of new
urns. In such a general framework, this model is associated to the
Yule-Simon distribution (Vitanov and Ausloos,  2012, 2015) 
whose density function $f$ is  \begin{equation} f(a;b) = b\,B(a,
b+1),\end{equation} \textbf{where $a$ is a positive integer, $b>0$,
}$B (a; b) $ is the Euler Beta function
\begin{equation}B(a; b) = \frac{ \Gamma(a) \Gamma(b)}{\Gamma(a+b)},\end{equation}
$\Gamma(x)$ being the standard Gamma function 
  (Abramowitz and  Stegun, 1970;  Gradshteyn and  Ryzhik, 2000).
  Explicitly,

\begin{equation}\label{Beta}
B(a,b) = \int_0^1 x^{a-1} (1-x)^{b-1}dx
\end{equation}
denotes the $Beta$-function; a random variable $X$  is
Beta-distributed if its probability density function (pdf)  obeys
 \begin{equation}\label{fx}
f(x)   = \frac{x^{a-1}\;(1-x)^{b-1}}{B(a,b)}.
\end{equation}


In practical words, newly created urn starts out with $k_0$ balls and further balls are added to urns at a rate proportional to the
number $k$ that they already have plus a constant $a\ge -k_0$.  With these definitions,   the fraction $P(k)$ of urns (areas)  having $k$ balls (cities)  in the limit of long time is given by
\begin{equation}
 P(k) = \frac{ B(k+a;b)}{B(k_0+a;b-1)}
 \end{equation} for $k\ge0$ (and zero otherwise).
In such a limit, the preferential attachment process generates a
long-tailed distribution following  a  hyperbolic (Pareto)
distribution, i.e. a power law, in  its tail.

\subsection{$K$ and $S$ parametrization}
\label{sub:beta} The relevant Beta-function moments, i.e.,
$K$ and $S$,  are given by  Johnson  and  Kotz (1970, pages 40-44), 
and recalled by Hanson (1991), 
 in terms of
$ a$ 
 and $ b$ 
parameters of the Beta function for the normalized variables:

 \begin{equation}
 K= 2(b-a)\frac{\sqrt{a+b+1}}{\sqrt{ab} (a+b+2)}
 \end{equation}\label{gamma3}

 \begin{equation}
 S = 3(a+b+1)\frac{2(a+b)^2 +ab (a+b-6)}{ab(a+b+2)(a+b+3)};
 \end{equation}\label{gamma4}
In order to develop the algebra, one also  introduces a so called
"help variable"  (Hanson, 1991)
\begin{equation}\label{help}
\rho =6\; \frac{K-S^2-1} {6+3S^2-2K}
\end{equation}
Notice that if Eq. (\ref{KpS2q}) holds, then
  \begin{equation}\label{helppq}
\rho =6\; \frac{(p-1)S^2+(q-1)} {(3-2p)S^2+2(3-q)},
\end{equation}
allowing a theoretical estimate at once, e.g.  if $p$
and $q$ are as in Table \ref{pqTableparameters}, and a possible
comparison to empirical results, shown on the last line in Table
\ref{Tablestat3}.
In fact, $ \rho = a+b$. 
 Thus,   one can obtain   $a$ and $b$ from:  

 \begin{equation}
ab=a(\rho-a)=\frac{6\rho^2(\rho+1)}{(r\rho+2)(\rho+3)K-3(\rho-6)(\rho+1)}
 \end{equation}\label{ab}

which leads to 2 solutions for $a$:
  \begin{equation}
a= \frac{\rho}{2} \left[ 1 \pm
\sqrt{1-\frac{24(\rho+1)}{(\rho+2)(\rho+3)K-3(\rho-6)(\rho+1)}}\right]
 \end{equation}

If the skewness  is positive then the larger solution will be the value of $b$
otherwise the larger solution will be the value of $a$  \textbf{(Hanson, 1991)}. 
The relevant values of the best fitted Beta can be read from Table
\ref{TableLav4parameters}.

%

\begin{figure} 
    \includegraphics[width=1.25\textwidth] {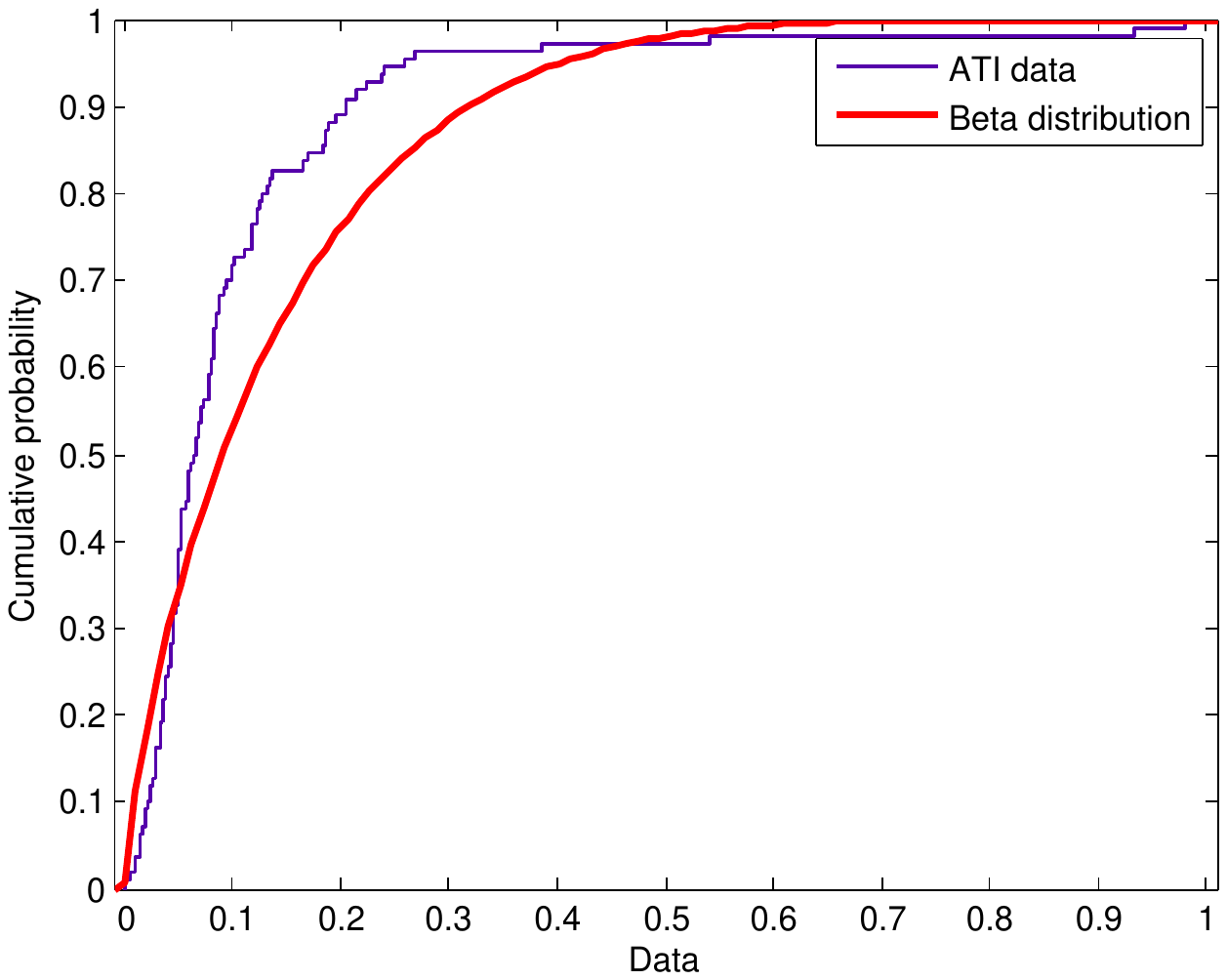}
\caption{\textbf{Cumulative distribution} function of the Beta
distribution, with parameters $a$ and $b$ calibrated according to
the ATI data for the IT provinces, in the reference year 2011,
\textbf{and empirical distribution of the data.} Fit parameters are
in Table \ref{pqTableparameters}.} \label{fig:BetaATI}
\end{figure}

\begin{figure} 
    \includegraphics[width=1.25\textwidth] {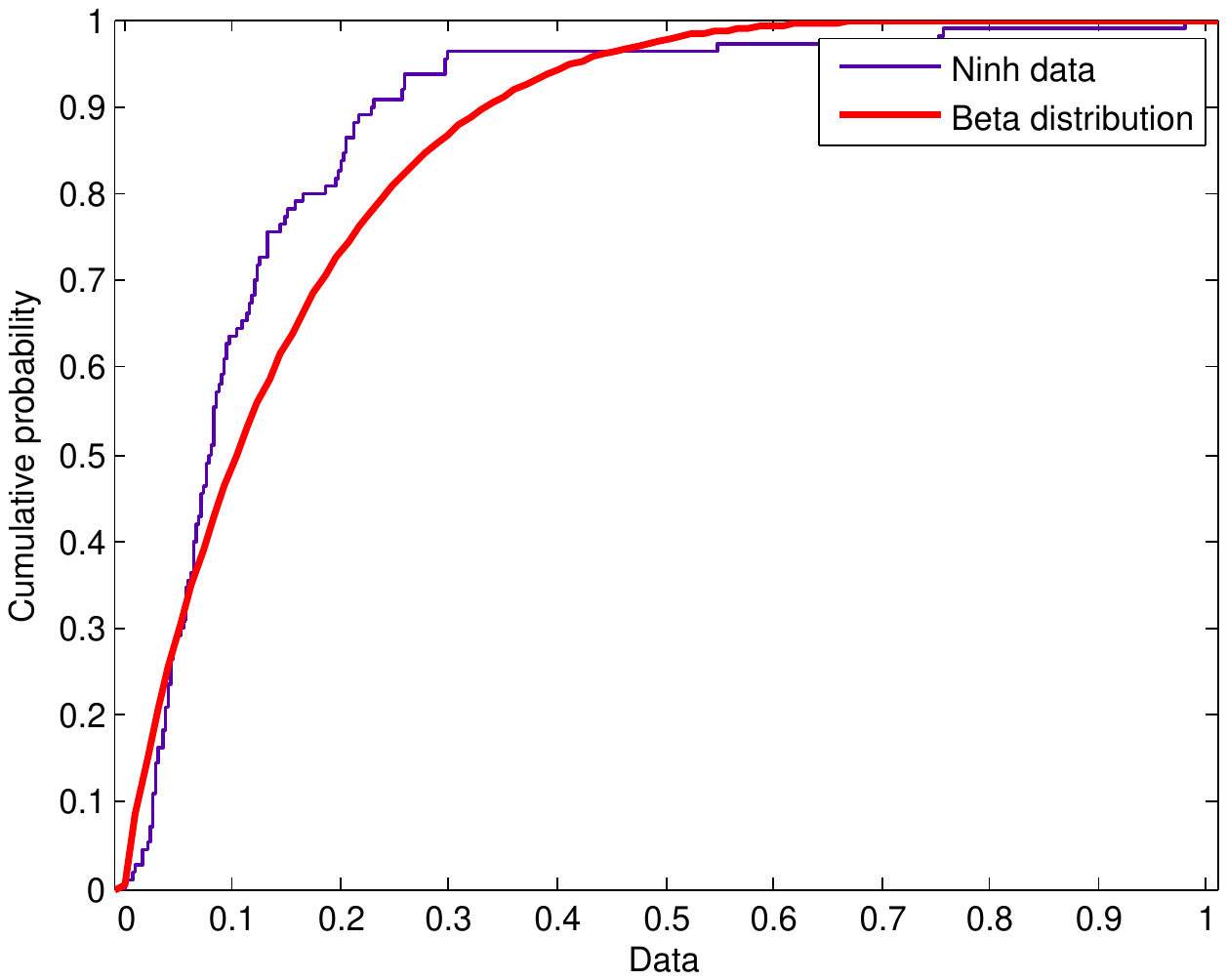}
\caption{\textbf{Cumulative distribution} function of the Beta
distribution, with parameters $a$ and $b$ calibrated according to
the demographic data $Ninhab$ for the IT provinces, in the reference
year 2011, \textbf{and empirical distribution of the data.} Fit
parameters are in Table \ref{pqTableparameters}.}
\label{fig:BetaNinhab}
\end{figure}

 Figures \ref  {fig:BetaATI} and \ref  {fig:BetaNinhab}
 illustrate the \textbf{cumulative distribution}
functions of the calibrated Beta distribution for the normalized ATI
and $N_{inhab}$ cases, respectively. We notice that the shapes of
the distributions are similar, and suggest a high concentration
around the small values of the Beta distribution.

\section{Conclusion} \label{conclusion}
 This paper explores the relationship between
skewness $S$ and kurtosis $K$ for series of demographical and
economical data. The considered sample is taken from the Italian
National Institute of Statistics (number of inhabitants) and the
Ministry of Economics and Finance (income taxes) for the Italian
cities, aggregated at a provincial level; the reference year is
2011. The existence of a quadratic and a power law relation between
skewness and kurtosis is illustrated, with fits which are both
visually quite appealing and markedly statistically sound. These findings support and add to
the empirical literature on the connections between $K$ and $S$;
 this is the first time that such quadratic and power laws
rules are found for socio-economic  surveys.

 It should seem interesting to search for the general conditions  leading to distributions with  such apparently simple relations.

In our  presently investigated  "socio-economic case", the empirical results have been supported by a theoretical argument
based on the Polya urn. In so doing, one gets the data-driven
calibration of the parameters of a Beta distribution, which adds
further insights on the different nature of economic and
demographic data.

\vskip0.3cm
\section*{Appendix A: Descriptive statistics}

Table \ref{TableProvATINinhabNc} lists   the ATI (in EUR) of each
provinces in IT, given in alphabetical order of their legal acronym,
the province population, and for general information, the number of
cities in the relevant province, in 2011.

In Table \ref{Tablestat3}, one displays a summary of  the (rounded)
statistical characteristics  for the distribution of  city ATI $
ATI_{c,p}$ of  each IT  provinces ($N_p=110$)   in 2011, for the
distribution of the number of inhabitants in a given province,
$N_{inhab,c}$  and for information the number of cities in the
relevant province, $N_{c,p}$.

In Table \ref{Tablestat3},  we also give $\sigma/\mu$, called the
coefficient of variation (CV), allowing to have some confidence  in
a relatively peaked distribution, -  if  CV is not too large. For
completeness, we also provide the immediately deduce  value of an
indirect measure, 3$(\mu-m)/\sigma$.

\begin{figure} 
    \includegraphics[width=1.250\textwidth]{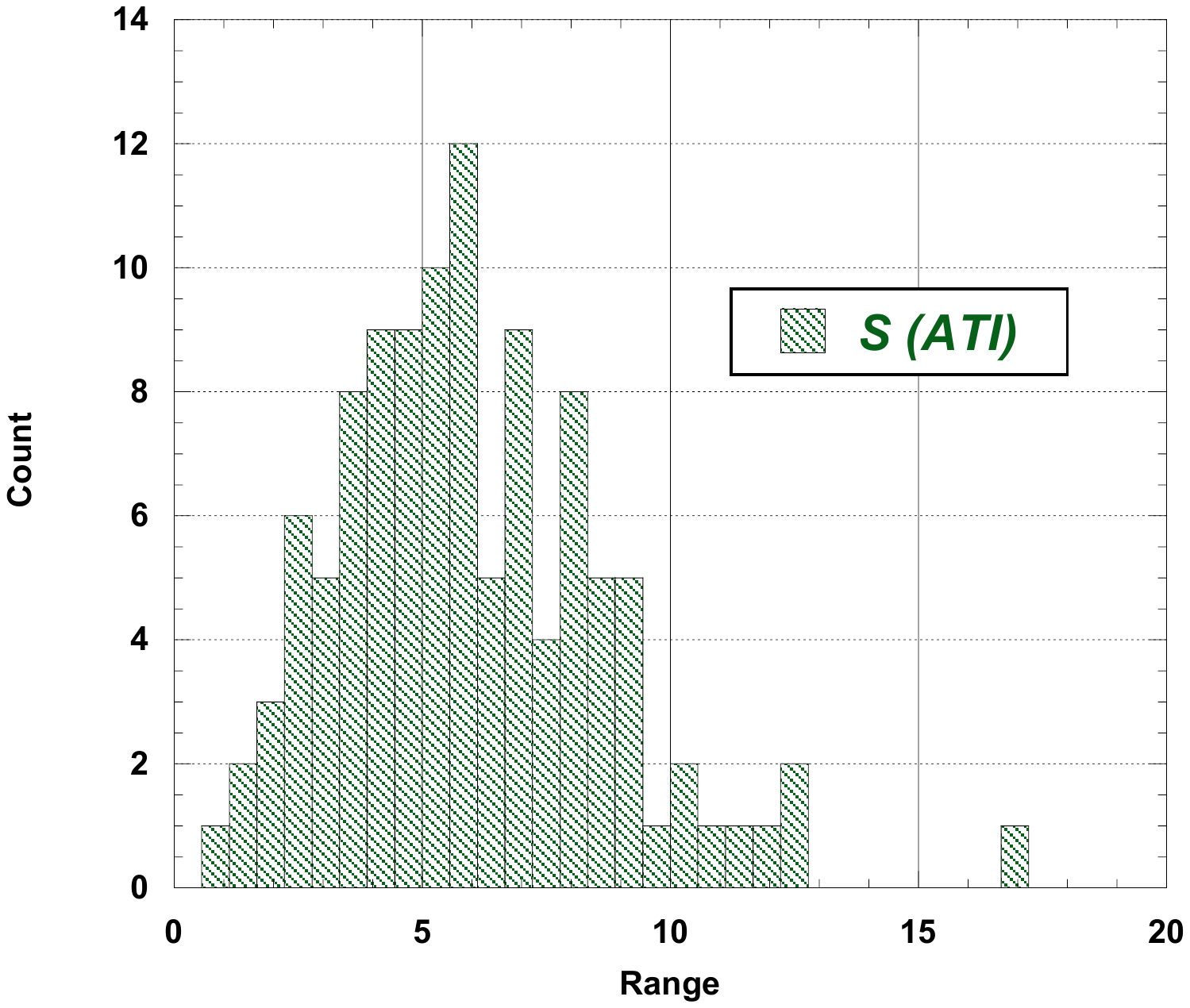}
    \caption{ Histogram  of the skewness of the ATI's
    for the reference year 2011 and for the IT provinces.}
   \label{fig:Plot2histoSATI}
\end{figure}

\begin{figure} 
    \includegraphics[width=1.250\textwidth]{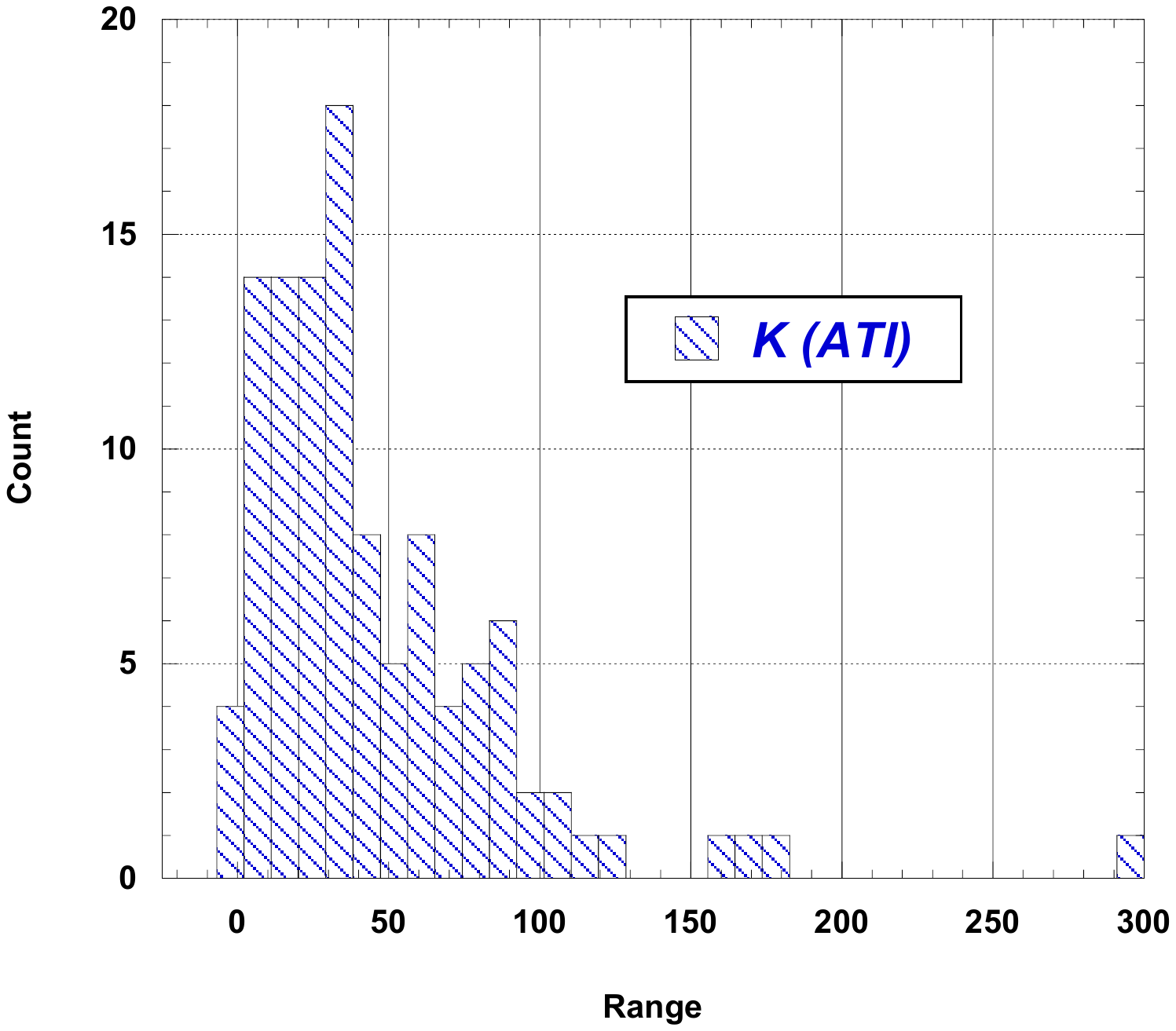}
    \caption{ Histogram  of the kurtosis of the ATI's
    for the reference year 2011 and for the IT provinces.}\label{fig:Plot3histoKATI}
\end{figure}

\begin{figure} 
    \includegraphics[width=1.250\textwidth]{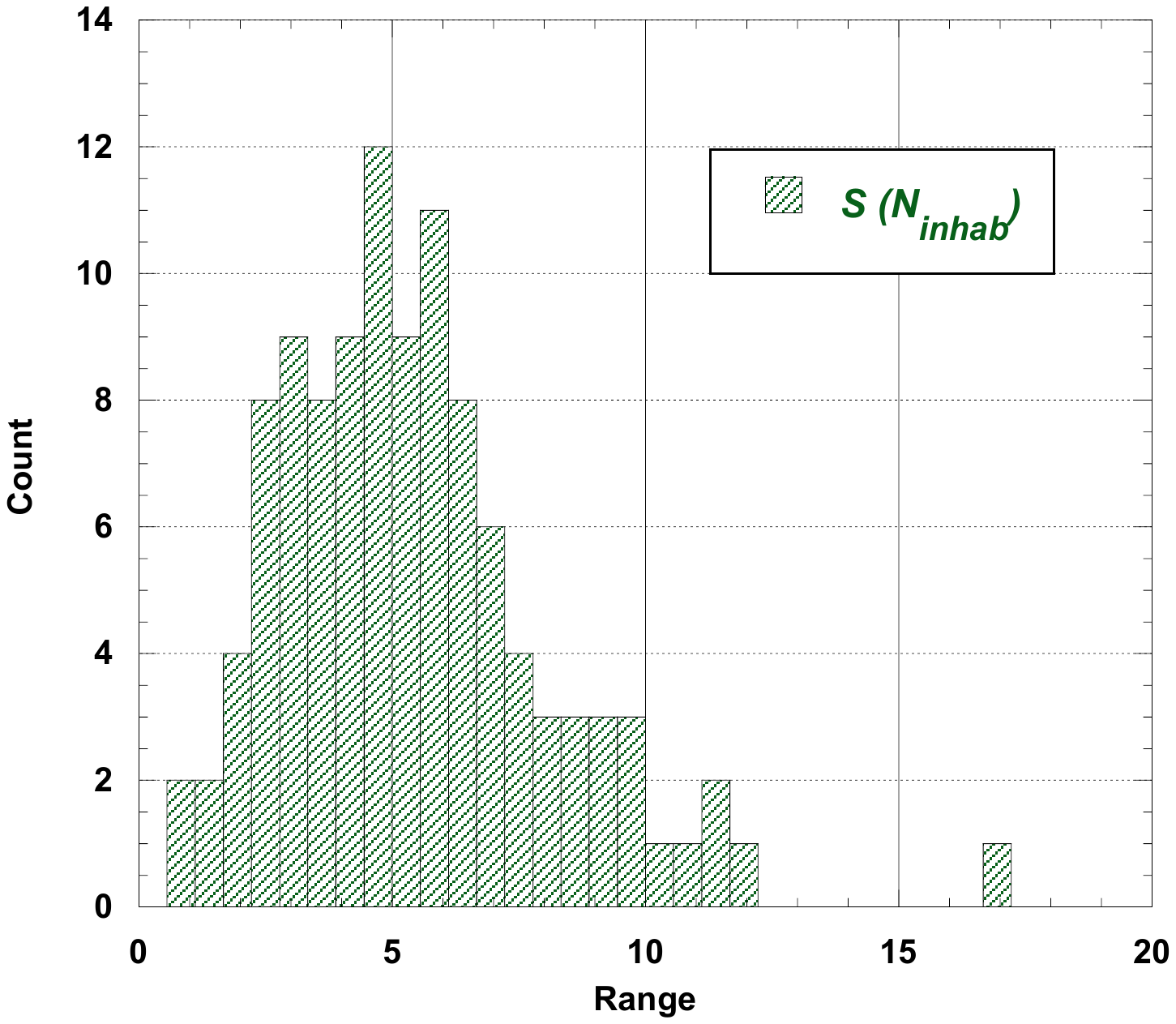}
    \caption{ Histogram  of the skewness of the     number of inhabitants
    for the reference year 2011 and for the IT provinces.}\label{fig:Plot4histoSNinhab}
\end{figure}

\begin{figure} 
    \includegraphics[width=1.250\textwidth]{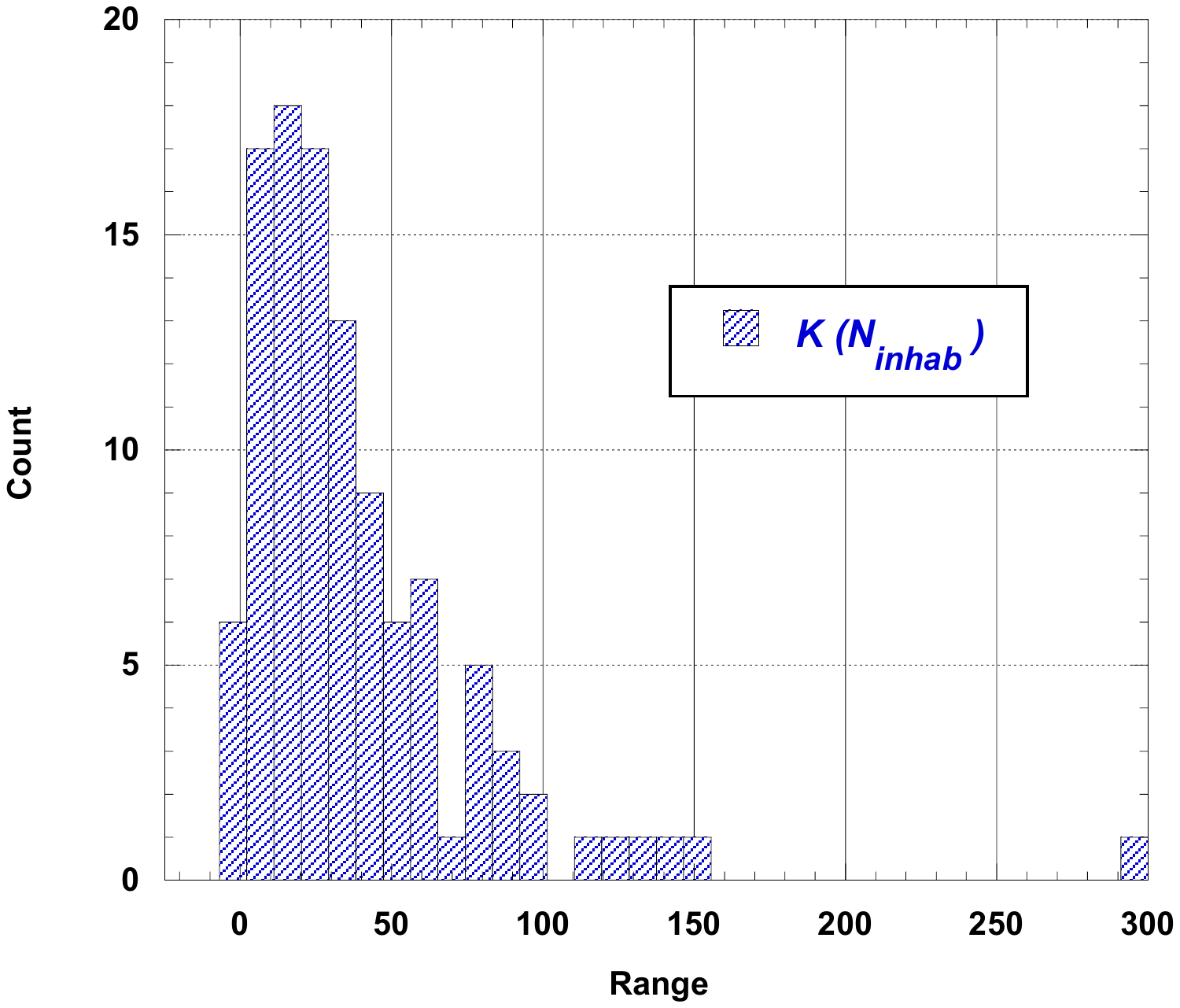}
    \caption{ Histogram  of the kurtosis of the
    number of inhabitants
    for the reference year 2011 and for the IT provinces.}\label{fig:Plot5histoKNinhab}
\end{figure}

Figures \ref{fig:Plot2histoSATI}-\ref{fig:Plot5histoKNinhab} provide
a  view of the empirical distributions of the skewness and kurtosis
for either the ATI or  the number of inhabitants of the IT
provinces. The shapes of the distributions for $S$ and for $K$ are
quite similar. In both cases some outliers emerge, \textbf{according
to the definition of outliers as those values outside the interval
$(\mu-2\sigma,\mu+2\sigma)$; see the values reported in Table
\ref{TablestatSKSK} and just compare them with the histograms in
Figures \ref{fig:Plot2histoSATI}-\ref{fig:Plot5histoKNinhab}.}

Notice that both distributions of $S$ and $K$ seem to be more
concentrated around their mean values for $N_{inhab}$ rather than
for $ATI$, hence suggesting a more evident regularity in the
asymmetry and in the peaks of the distribution across provinces in
the former case. \textbf{Moreover, Figures
\ref{fig:Plot2histoSATI}-\ref{fig:Plot5histoKNinhab} show also that
skewness and kurtosis of our specific dataset, unexpectedly, seem to
depart from a normal asymptotic distribution. }

\clearpage

 \begin{table}
  \begin{center} \begin{tabular}{|c|c|c|c|c|c|c|c|c|c|c|c|c|c|c|c|}
\hline prov  &    $ATI$ & Ninhab&$N_{c,p}$ &prov &$ATI$&
Ninhab& $N_{c,p}$   &prov  &$ATI$&  Ninhab& $N_{c,p}$         \\
\hline
 AG &   2.844722    &   447310  &   43  &   FR  &   4.488987    &   493928  &   91  &   PU  &   4.235173    &   363003  &   60  &   \\
AL  &   5.738100    &   428417  &   190 &   GE  &   13.45363    &   862267  &   67  &   PV  &   7.602668    &   537620  &   190 &   \\
AN  &   6.176045    &   475038  &   49  &   GO  &   1.973022    &   139983  &   25  &   PZ  &   3.137922    &   378409  &   100 &   \\
AO  &   1.873498    &   126982  &   74  &   GR  &   2.651100    &   221442  &   28  &   RA  &   5.322050    &   384575  &   18  &   \\
AP  &   2.286054    &   209887  &   33  &   IM  &   2.422774    &   212854  &   67  &   RC  &   4.151080    &   547897  &   97  &   \\
AQ  &   3.149092    &   297418  &   108 &   IS  &   0.808795 &  87578   &   52  &   RE  &   7.297694    &   518011  &   45  &   \\
AR  &   4.227090    &   344453  &   39  &   KR  &   1.030377    &   171331  &   27  &   RG  &   2.210240    &   308329  &   12  &   \\
AT  &   2.709870    &   217870  &   118 &   LC  &   5.033768    &   336705  &   90  &   RI  &   1.681281    &   156142  &   73  &   \\
AV  &   3.343223    &   430292  &   119 &   LE  &   6.149541    &   803554  &   97  &   RM  &   59.68562    &   4042676 &   121 &   \\
BA  &   11.31215 &  1248086 &   41  &   LI  &   4.414274    &   336412  &   20  &   RN  &   3.812157    &   322294  &   27  &   \\
BG  &   14.54681 &  1087401 &   244 &   LO  &   3.102357    &   224393  &   61  &   RO  &   2.871922    &   242409  &   50  &   \\
BI  &   2.540291    &   182417  &   82  &   LT  &   5.154685    &   544391  &   33  &   SA  &   8.297128    &   1091227 &   158 &   \\
BL  &   2.798485    &   210277  &   69  &   LU  &   4.860075    &   388922  &   35  &   SI  &   3.740127    &   267194  &   36  &   \\
BN  &   2.134726    &   285677  &   78  &   MB  &   12.98964    &   841102  &   55  &   SO  &   2.211207    &   181091  &   78  &   \\
BO  &   16.17945 &  981807  &   60  &   MC  &   3.615307    &   319181  &   57  &   SP  &   2.968193    &   220063  &   32  &   \\
BR  &   3.225014    &   401207  &   20  &   ME  &   5.437395    &   653470  &   108 &   SR  &   3.196919    &   397952  &   21  &   \\
BS  &   15.63700 &  1240553 &   206 &   MI  &   55.71135 &  3072152 &   134 &   SS  &   3.303827    &   329616  &   66  &   \\
BT  &   2.536561    &   391127  &   10  &   MN  &   5.256847    &   408893  &   70  &   SV  &   3.797833    &   282255  &   69  &   \\
BZ  &   7.469868    &   505067  &   116 &   MO  &   9.965410    &   687237  &   47  &   TA  &   5.131752    &   579836  &   29  &   \\
CA  &   5.858126    &   552303  &   71  &   MS  &   2.388878    &   200387  &   17  &   TE  &   2.907707    &   305872  &   47  &   \\
CB  &   0.200671    &   226982  &   84  &   MT  &   1.637990    &   200842  &   31  &   TN  &   7.262721    &   526510  &   217 &   \\
CE  &   6.270454    &   906600  &   104 &   NA  &   23.17133    &   3058592 &   92  &   TO  &   32.29885    &   2245252 &   315 &   \\
CH  &   3.758159    &   388280  &   104 &   NO  &   5.194036    &   364217  &   88  &   TP  &   3.187134    &   430843  &   24  &   \\
CI  &   1.111378    &   128581  &   23  &   NU  &   1.318428    &   158456  &   52  &   TR  &   2.742176    &   228944  &   33  &   \\
CL  &   1.871670    &   273155  &   22  &   OG  &   0.457826 &  57492   &   23  &   TS  &   3.709890    &   233077  &   6   &   \\
CN  &   7.447040    &   586599  &   250 &   OR  &   1.331566    &   164113  &   88  &   TV  &   11.37026    &   877905  &   95  &   \\
CO  &   8.109499    &   587547  &   160 &   OT  &   1.496835    &   151627  &   26  &   UD  &   7.305619    &   536035  &   136 &   \\
CR  &   4.857371    &   357473  &   115 &   PA  &   10.25093 &  1239837 &   82  &   VA  &   12.30706    &   873241  &   141 &   \\
CS  &   5.017154    &   715485  &   155 &   PC  &   4.073974    &   284711  &   48  &   VB  &   1.953740    &   160385  &   77  &   \\
CT  &   8.251406    &   1080034 &   58  &   PD  &   12.33167 &  921659  &   104 &   VC  &   2.381702    &   176853  &   86  &   \\
CZ  &   2.839421    &   360165  &   80  &   PE  &   3.319864    &   315629  &   46  &   VE  &   11.29438 &  850523  &   44  &   \\
EN  &   1.157228    &   173668  &   20  &   PG  &   7.750671    &   657535  &   59  &   VI  &   11.06162 &  859987  &   121 &   \\
FC  &   5.071875    &   390381  &   30  &   PI  &   5.434720    &   412729  &   39  &   VR  &   11.77778    &   903564  &   98  &   \\
FE  &   4.787859    &   353725  &   26  &   PN  &   4.150770    &   310983  &   51  &   VS  &   0..342572 & 101396  &   28  &   \\
FG  &   4.562346    &   627007  &   61  &   PO  &   3.101017    &   246219  &   7   &   VT  &   3.264761    &   313998  &   60  &   \\
FI  &   14.30128 &  971437  &   44  &   PR  &   6.737250    &   428652  &   47  &   VV  &   1.119749    &   161952  &   50  &   \\
FM  &   1.793660    &   175047  &   40  &   PT  &   3.368717    &   288415  &   22  &                                   \\

 \hline
\end{tabular}  \end{center}
\caption{ Summary table of the ATI (in EUR), number
of inhabitants and number of cities of provinces in IT (in e+09
units) in the reference year 2011. Acronyms for provinces are those
legally used,  in IT.}  \label{TableProvATINinhabNc}
  \end{table}

   \begin{table} \begin{center}
\begin{tabular}[t]{|cc||cc||cc|}
  \hline
    \multicolumn{2}{|c|}{ $ATI_{c,p}$}&  \multicolumn{2}{|c|}{ $N_{inhab,c,p}$}  & \multicolumn{2}{|c|}{ $N_{c,p}$} \\ \hline
min.(x$10^{-5}$)    &   3.3479  &min.(x$10^{-4}$)&  5.7492  &   Min.    &   6   \\
Max.(x$10^{-10}$)   &   4.5490  &Max.(x$10^{-6}$)   &   4.0276  &   Max.    &   315 \\
Sum(x$10^{-11}$)    &   7.2184  &Sum(x$10^{-7}$)&   5.9571 &    Sum &   8092    \\
mean($\mu$)(x$10^{-7}$)     &   8.9204  &mean($\mu$)(x$10^{-5}$)&   5.4155 &    Mean    &   73.564  \\
median($m$)(x$10^{-7}$)     &   2.4601  &median($m$)(x$10^{-5}$)&   3.7131 &    Median  &   60  \\
RMS(x$10^{-8}$)     &   6.7701  &RMS(x$10^{-5}$)    &   7.9654 &    RMS &   91.902  \\
Std.Dev.($\sigma$)(x$10^{-8}$)  &   6.7115  &Std.Dev.($\sigma$)(x$10^{-5}$) &   5.8680 &    Std.Dev.    &   55.338  \\Var.(x$10^{-17}$)     &   4.5044  &Var.(x$10^{-11}$) &    3.4433  &   Var.    &   3062.27 \\
Std.Err.(x$10^{-6}$)    &   7.4609  &   Std.Err.(x$10^{-6}$)    &   0.5595  &   Std.Err.    &   5.2762  \\
Skewn.  &   49.490  &   Skewn.  &   3.6571  &   Skewn.  &   1.7294  \\
Kurt.   &   2994.7  &   Kurt.   &   15.873  &   Kurt.   &   3.6845  \\\hline
$\mu/\sigma$    &   0.1329  &   $\mu/\sigma$    &   0.9229  &   $\mu/\sigma$    &   1.3294  \\
3$(\mu-m)/\sigma$   &   0.2889  &3$(\mu-m)/\sigma$  &   0.8703  &   3$(\mu-m)/\sigma$   &   0.7353  \\
$r$ &   2.4020  &   $r$     &   0.6254  &   $r$     &   -0.2417 \\\hline
\end{tabular}
   \caption{Summary of  (rounded) statistical
characteristics  for  the distribution of ATI of cities   in IT
provinces,   for the  distribution of inhabitants in IT cities in
the various ($N_p=110$) provinces, and   for the distribution of the
number of cities in provinces    in 2011.
}\label{Tablestat3}
\end{center} \end{table}

\clearpage
\clearpage

\clearpage

\clearpage

\end{document}